\documentclass[aps,prd,groupedaddress,graphicx,nofootinbib]{revtex4}
\usepackage{amsmath,amssymb,graphics,graphicx,color}

\begin{document}

\title{Towards an analytic solution of rapid roll inflation with a quartic potential}
%\author{Kazunori Kohri and Chia-Min Lin}

%\author{Kazunori Kohri$^{1}$}\email{kohri@post.kek.jp}
%\author{C. S. Lim$^{2}$}\email{lim@lab.twcu.ac.jp}
\author{Chia-Min Lin}\email{cmlin@ncut.edu.tw}
\affiliation{Fundamental Education Center, National Chin-Yi University of Technology, Taichung 41170, Taiwan}

%\affiliation{$^1$Cosmophysics Group, Theory Center, IPNS
%KEK, and The Graduate University for Advanced Studies (Sokendai), 1-1
%Oho, Tsukuba 305-0801, Japan}
%\affiliation{$^{2,3}$Department of Physics, Kobe University, Kobe 657-8501, Japan}
%\affiliation{$^2$Department of Mathematics, Tokyo Woman's Christian University, Tokyo 167-8585, Japan}
%\affiliation{$^3$Department of Physics, Chuo University, Bunkyo-ku, Tokyo 112, Japan}

%\baselineskip 14pt

\date{Draft \today}

\begin{abstract}
In this paper, I present an analytic solution to the equation of motion of the inflaton field in a model of rapid roll inflation with a quartic potential by assuming that the Hubble parameter is a constant during inflation. The result is obtained by using Jacobi elliptic functions. The spectral index $n_s$ and the running spectral index $n_s^\prime$ is obtained without using rapid roll or extended slow roll approximations. The cosmological consequences of the model is discussed.
\end{abstract}
\maketitle

\section{Introduction}

\large
\baselineskip 18pt

Slow roll inflation requires that the effective inflaton mass should be much smaller than the Hubble parameter during inflation, that is $m^2 \ll H^2$. However, when we try to build an inflation model in the framework of string theory or supergravity, one often encounter an Hubble induced inflaton mass, that is $m^2 \sim H^2$ \cite{Copeland:1994vg, Kachru:2003sx}. In addition, scalar field $\phi$ conformally coupled to gravity acquires effective mass term $\frac{1}{2}\xi R \phi^2$ with $\xi=\frac{1}{6}$ and the effective mass $m^2 \sim 12\xi H^2=2H^2$ which violates slow-roll condition. However, inflation can still happen in the form of rapid roll inflation \cite{Kofman:2007tr}.

In order to study rapid roll inflation, rapid roll conditions or extended slow roll conditions are used as an approximation to deal with the problem \cite{Kofman:2007tr, Chiba:2008ia, Kobayashi:2008rx, Kobayashi:2009nv}. In this paper, I analyze a specific (but quite representative) case by considering a quartic term in the potential without using these approximation methods. In the following, I use the system of units $M_P=2.4 \times 10^{18}\mbox{ GeV}=(8 \pi G)^{-1/2}=1$. 

\section{Inflation with a quadratic potential}
As a warm up, I will review a simple inflation model with the potential of the inflaton field $\phi$ as
\begin{equation}
V=V_0+\frac{1}{2}m^2 \phi^2,
\end{equation}
where the constant $V_0$ during inflation can be produced by the mechanism of hybrid inflation \cite{Linde:1993cn}. 
The equation of motion is given by
\begin{equation}
\ddot{\phi}+3H\dot{\phi}+m^2\phi=0.
\label{eq2}
\end{equation}
Slow roll approximation assumes the first term on the left-hand side of the equation is negligible, therefore the number of e-folds is given by
\begin{equation}
N \equiv \int H dt = \int \frac{H}{\dot{\phi}}d \phi = -\int \frac{3H^2}{m^2 \phi}d \phi = -\frac{3 H^2}{m^2} \ln \phi +C=Ht,
\end{equation}
where $C$ is an integration constant and it is assumed that the Hubble parameter is a constant during inflation.
Therefore by defining $\phi(0)=\phi_0$ we can write 
\begin{equation}
\phi=\phi_0 e^{-\frac{m^2}{3H^2}N}=\phi_0 e^{-\frac{m^2}{3H^2}Ht}.
\label{eq4}
\end{equation}
Actually in this case, we can do better than imposing slow-roll approximation.
With the assumption that the Hubble parameter is a constant during inflation, we can solve Eq.~(\ref{eq2}) by substituting the ansatz $\phi=Be^{bt}$ into the equation and obtain%\footnote{It should be clear from the context that here $a$ is not the scale factor $a(t)$ in the metric.}
\begin{equation}
b=\frac{-3H \pm 3H\sqrt{1-\frac{4m^2}{9H^2}}}{2}.
\end{equation}
We can neglect the solution with the minus sign because it decays faster, and the result is
\begin{equation}
\phi=\phi_0 e^{\frac{-3H + 3H\sqrt{1-\frac{4m^2}{9H^2}}}{2}t}.
\label{eq6}
\end{equation}
When $m^2 \ll H^2$, Eq.~(\ref{eq6}) reduces to 
\begin{equation}
\phi=\phi_0 e^{\frac{-3H + 3H\left(1-\frac{2m^2}{9H^2}\right)}{2}t} = \phi_0 e^{-\frac{m^2}{3H^2}Ht},
\end{equation}
which recovers Eq.~(\ref{eq4}) and shows the validity of slow-roll approximation.

\section{Rapid Roll Inflation}
\label{sec1}
The action of a non-minimally coupled scalar field $\phi$ is given by
\begin{equation}
S=\int d^4x \sqrt{-g}\left[\frac{R}{2}-\frac{1}{2}\partial ^\mu \phi \partial_\mu \phi -V(\phi)-\frac{1}{2}\xi R \phi^2\right].
\end{equation}
In a FRW universe with $ds^2=-dt^2+a(t)^2 d \mathbf{x}^2$, the  equation of motion of the scalar field $\phi$ is
\begin{equation}
\ddot{\phi}+3H\dot{\phi}+6\xi (\dot{H}+2H^2)\phi+V^\prime (\phi)=0,\label{eq8},  
\end{equation}
and the Friedmann equation is
\begin{equation}
3H^2=\frac{1}{2}(\dot{\phi}+H\phi)^2+V(\phi) \equiv \frac{1}{2}\pi^2+V(\phi).\label{eq9}
\end{equation}
If we assume $H$ and $V=V_0$ are constants during inflation and choosing conformal coupling $\xi=\frac{1}{6}$, Eq.~(\ref{eq8}) becomes
\begin{equation}
\ddot{\phi}+3H\dot{\phi}+2H^2\phi=0.
\end{equation}
This is nothing but Eq.~(\ref{eq2}) with $m^2=2H^2$ and the solution can be obtained from Eq.~(\ref{eq6}) as
\begin{equation}
\phi = \phi_0 e^{-Ht}.
\label{phis}
\end{equation} 
If we make a time derivative,
\begin{equation}
\dot{\phi}=-H\phi,
\label{phidot}
\end{equation}
and substitute the result into Eq.~(\ref{eq9}), we obtain 
\begin{equation}
3H^2=V_0,
\end{equation}
which is indeed a constant and implies a de Sitter universe. This also shows the important role played by $V_0$. 
\section{rapid roll inflation with a quartic potential}
\label{sec2}
Now let us go beyond the simplest model and consider a potential of the form\footnote{This potential was considered in the case of hilltop inflation where the parameter space is analyzed in the framework of slow-roll inflation \cite{Kohri:2007gq}.}
\begin{equation}
V=V_0+\frac{1}{2}M^2\phi^2+\frac{1}{4}\lambda \phi^4.
\end{equation}
This might be regarded as a ``two term approximation'' to a more general potential. 
By our assumption that Hubble parameter is a constant and conformal coupling $\xi=\frac{1}{6}$, Eq.~(\ref{eq8}) becomes
\begin{equation}
\ddot{\phi}+3H\dot{\phi}+(2H^2+M^2)\phi+\lambda\phi^3=0.
\end{equation}
I will ignore the quardratic term of the potential from now on by considering $M^2 \ll 2 H^2$ or $M^2=0$ since conformal coupling already gives $\phi$ a large effective mass\footnote{For example, if $M \sim 0.1 H$, we have $M^2 \sim 0.01 H^2$.}. Therefore the equation of motion becomes
\begin{equation}
\ddot{\phi}+3H\dot{\phi}+m^2\phi+\lambda\phi^3=0,
\label{eq15}
\end{equation}
and let us keep in mind that 
\begin{equation}
m^2 \equiv 2H^2.
\label{eq16}
\end{equation}
Is there an analytic solution to Eq.~(\ref{eq15})? This non-linear second-order equation is called a Duffing equation and corresponds to the equation of motion of a damped Duffing oscillator without a driving force. In order to solve the equation, let us consider the ansatz \cite{duf}
\begin{equation}
\phi(t)=\alpha(t) \mbox{cn} (\omega (t),k^2).
\end{equation}
Here $\mbox{cn}$ is a Jacobi elliptic function defined as 
\begin{equation}
\mbox{cn}(u,k^2)=\cos \psi \mbox{, where } u=\int_0^{\psi} \frac{d\theta}{\sqrt{1-k^2 \sin^2 \theta}}.
\end{equation}
By substituting our ansatz into Eq.~(\ref{eq15}), and use some properties of Jacobi elliptic functions (see the appendix), we could obtain
\begin{eqnarray}
\mbox{cn} (\omega (t),k^2)[m^2 \alpha+3 H \dot{\alpha}-\alpha \dot{\omega}^2+2k^2 \alpha\dot{\omega}^2+\ddot{\alpha}]+\mbox{cn} (\omega (t),k^2)^3[\lambda \alpha^3-2k^2 \alpha\dot{\omega}^2] \nonumber \\ -\mbox{sn} (\omega (t),k^2)\mbox{dn} (\omega (t),k^2)[3H \alpha\dot{\omega}+2\dot{\alpha}\dot{\omega}+ \alpha\ddot{\omega}]=0.
\end{eqnarray}
The above equation holds for all time $t$ if and only if
\begin{eqnarray}
m^2 \alpha+3 H \dot{\alpha}-\alpha \dot{\omega}^2+2k^2 \alpha\dot{\omega}^2+\ddot{\alpha}=0 \label{eq19}  \\
\lambda \alpha^3-2k^2 \alpha\dot{\omega}^2=0  \label{eq20}\\
3H \alpha\dot{\omega}+2\dot{\alpha}\dot{\omega}+\alpha\ddot{\omega}=0 \label{eq21}
\end{eqnarray}
Let us substitute the ansatz $\alpha=Ae^{\beta t}$ and $\omega = \Omega e^{\beta t}$ into Eq.~(\ref{eq21}), and obtain $\beta=-H$. Then after substituting this into Eq.~(\ref{eq19}), we can see that it can be satisfied if $k^2=1/2$ and $m^2=2H^2$. Interestingly, the latter condition is exactly Eq.~(\ref{eq16})!\footnote{Here this condition comes to us so that the term of coefficient $m^2$ cancels out with the term of coefficient $2H^2$ so that our ansatz can work. It may be interesting to investigate whether this ``miraculous'' cancellation is just a coincidence or there is some deep mathematical reason. } Finally from Eq.~(\ref{eq20}), we obtain $\Omega^2=\lambda A^2/H^2$, therefore $\Omega=\pm  \sqrt{\lambda} A/H$. It does not matter whether we take the plus or minus sign for $\Omega$ because $\mbox{cn}$ is an even function and we will take the plus sign. Actually there will be an integration constant $C$ for $\omega$ since it appears with at least one time derivative in the equations, so eventaully we have our analytical expression
\begin{equation}
\phi(t)=A e^{-Ht}\mbox{cn}\left(\frac{\sqrt{\lambda}A}{H}e^{-Ht}+C,\frac{1}{2} \right).
\label{phi}
\end{equation}
The time derivative of $\phi$ is given by using the equations in the appendix as
\begin{equation}
\dot{\phi}=-H\phi+\sqrt{\frac{\lambda}{2}}\sqrt{A^4e^{-4Ht}-\phi^4}.
\label{eq24}
\end{equation}
From the above two equations, the constants $A$ and $C$ can be determined by the initial condition $\phi(0)=\phi_0$ and $\dot{\phi}(0)=\dot{\phi}_0$ as

\begin{eqnarray}
A^4=\frac{\lambda \phi_0^4+2\dot{\phi}_0^2+4H\phi_0\dot{\phi}_0+2H^2\phi_0^2}{\lambda}\\
C=\mbox{cn}^{-1}\left(\frac{\phi_0}{A},\frac{1}{2}\right)-\frac{\sqrt{\lambda} A}{H}.
\end{eqnarray}

Now let us see what we can learn from the above solution.
From Eq.~(\ref{eq24}), we can obtain
\begin{equation}
\pi \equiv \dot{\phi}+H\phi=\sqrt{\frac{\lambda}{2}}\sqrt{A^4e^{-4Ht}-\phi^4}.
\label{pi}
\end{equation}
Therefore 
\begin{equation}
-\frac{\dot{H}}{H^2}=\frac{\pi^2+V^\prime \phi/2}{\pi^2/2+V}=\frac{\frac{\lambda}{2}A^4 e^{-4Ht}}{\frac{\lambda}{4}A^4 e^{-4Ht}+V_0},
\label{eq28}
\end{equation}
where Eqs.~(\ref{eq8}) and (\ref{eq9}) is used.
Our approximation that the Hubble expansion rate $H$ is a constant is valid if $|\dot{H}/H^2| \ll 1$.

From Eqs.~(\ref{pi}) and (\ref{phi}), we can obtain 
\begin{equation}
\pi = \sqrt{\frac{\lambda}{2}}A^2 e^{-2Ht}\sqrt{1-\mbox{cn}^4 \left( \frac{\sqrt{\lambda}A}{H}e^{-Ht}+C, \frac{1}{2} \right)}.
\end{equation}
This means for any (reasonable\footnote{For example, we do not choose the initial condition at the vaccum so that inflation does not happen.}) choices of the initial condition $\phi_0$ and $\dot{\phi}_0$, which determines $A$ and $C$, $\pi$ goes exponentially fast towards the attractor $\pi \sim 0$, namely $\dot{\phi} \sim -H \phi$, because the Jacobi ellipic function $\mbox{cn}$ is bounded.

\section{Primordial curvature perturbation}
\label{dpfmr}

Since in rapid roll inflation, the inflaton field is rolling rapidly, primordial density perturbation cannot come from the fluctuation of the inflaton field. We can consider a curvaton scenario\cite{Enqvist:2001zp, Lyth:2001nq, Moroi:2001ct}, modulated reheating scenatio\cite{Dvali:2003em,Kofman:2003nx}, or even a curvaton with modulated decay\cite{Langlois:2013dh, Assadullahi:2013ey, Kohri:2013soz}. The following conclusion are basically the same.
As an illustration, let us consider a modulated reheating scenario, where the primordial curvature perturbation $\zeta$ is given by the modulated decay width $\Gamma$ through a modulated coupling constant $\Lambda$ from a fluctuating scalar field $\chi$ as 
\begin{equation}
\zeta \sim \frac{\delta \Gamma}{\Gamma} \propto \frac{\delta \Lambda}{\Lambda} \propto \frac{\delta \chi}{\chi},
\end{equation}
where $\Gamma \sim \Lambda^2 M$. The idea is motivated because in string theory, there is ``no free parameters'' in the sense that the ``coupling constants'' are determined dynamically by scalar fields. Therefore the quantum fluctuations of a light scalar field resulted in the fluctuation of the coupling constant, hence the decay width.
At horizon crossing, $k=aH=a_0He^{\int H dt}$.
Since $\delta \chi \sim H$, \textit{if we assume the effective mass of $\chi$ is much smaller than the Hubble parameter}, the tilt of the spectrum can be expressed as 
\begin{equation}
n_s-1=\frac{d \ln H^2}{d \ln k } = \frac{d \ln H^2}{d t} \frac{dt}{d \ln \left(a_0He^{\int H dt}\right)} = 2\frac{\dot{H}}{H^2}\left(1+\frac{\dot{H}}{H^2} \right)^{-1},
\label{eq30}
\end{equation} 
and the running spectral index is given by
\begin{equation}
n_s^\prime \equiv \frac{d n_s}{d \ln k}=2 \left( 1+\frac{\dot{H}}{H^2} \right)^{-3} \frac{1}{H} \left( \frac{\dot{H}}{H^2} \right)^{\cdot}.
\end{equation}
From Eq.~(\ref{eq28}), we can obtain 
\begin{equation}
n_s-1=-\frac{\lambda A^4 e^{-4Ht}}{V_0-\frac{\lambda}{4}A^4 e^{-4Ht}},
\end{equation}
and 
\begin{equation}
n_s^\prime =\frac{\lambda A^4 e^{-4Ht}}{V_0-\frac{\lambda}{4}A^4 e^{-4Ht}} \cdot \frac{4 V_0 \left( \frac{\lambda}{4}A^4 e^{-4Ht}+V_0 \right)}{\left( V_0-\frac{\lambda}{4}A^4 e^{-4Ht} \right)^2  } \simeq -4(n_s-1).
\end{equation}
The approximation used above is not indispensable, because given $n_s$ we can calculate the ratio $\lambda A^4 e^{-4Ht} / V_0$ and the result can be used to obtain $n_s^\prime$.
However, this result is inconsistant with the latest Planck data \cite{Akrami:2018odb} which gives $n_s-1 \simeq -0.04$ and $|n_s^\prime | \lesssim 0.01$. Therefore our assumption to ignore the mass of $\chi$ may be wrong. If the effective mass of $\chi$ is non-negligible, there is a correction to the spectral tilt as
\begin{equation}
\Delta n_s = \frac{2}{3}\frac{V^{\prime\prime}(\chi)}{H^2}.
\end{equation}
We can fit Planck data $|n_s^\prime | \lesssim 0.01$ if 
\begin{equation}
\frac{\lambda A^4 e^{-4Ht}}{V_0} \lesssim 0.0025
\end{equation}
at horizon exit and obtain $n_s-1 \simeq -0.04$ if $V^{\prime\prime}(\chi) \sim -0.06H^2$. This implies the effective mass of $\chi$ is about the same order of magnitude as the Hubble parameter. Since in this case the potential of $\chi$ has a hilltop form, we may call $\chi$ a ``hilltop modulon''. The same result can also be achieved by considering a hilltop curvaton \cite{Matsuda:2007av}.
  
Instead of considering a hilltop modulon or hilltop curvaton, another possibility to evade Planck constraint is to have $n_s \sim 1$. This might be achieved if some part of primordial density perturbation comes from cosmic string contribution \cite{Bevis:2007gh, Battye:2007si}. Since $V_0$ presumably comes from the mechanism of hybrid inflation, the production of cosmic string is quite generic \cite{Jeannerot:2003qv}. However, without further simulations by using the latest observational results, it is not clear whether $n_s \sim 1$ can still be achieved by considering cosmic string contribution.

%Therefore by including a quartic term in the postential with a positive $\lambda$, we have $\dot{H}/H^2 < 0$, which implies a red spectrum. The simplest rapid roll inflation which corresponds to $\lambda=0$ resulted in a scale-independent Harrison-Zeldovich (HZ) spectrum $n_s=1$, which is decisively ruled out by Planck \cite{Akrami:2018odb}. Therefore in this framework, it is necessary to go beyond the simplest model where only an (induced) quadratic mass term is included. To include a quartic potential term is interesting in its own right in the sense of conformal invariance, and it can also be regarded as a Taylor expansion approximation to many more complicated potential.

\section{Conclusion and Discussion}
\label{con}

In this paper, I have presented an analytic expression of solutions for a rapid roll inflation with a quartic term in the potential. Although nowadays numerical solution is readily available, it is still satisfying if an analytic expression can be found even for simpler cases, and it could shed some light to improve our understanding of rapid roll inflation.

In our model, there are free parameters $\lambda$ and $V_0$. We may realize this model in the framework of supersymmetric particle physics theories along the line of modified supernature inflation \cite{Lin:2009yt, Kohri:2010sj, Kohri:2013gva}, where $V_0$ is determined by SUSY breaking scale and the dimensionless parameter $\lambda$ is given by the ratio between a soft SUSY breaking parameter and the Planck scale, therefore it could be naturally small. I will leave the investigation to future work.

\appendix
\section{Jacobi Elliptic Functions}
The Jacobi elliptic function $\mbox{cn}$ is defined as
\begin{equation}
\mbox{cn}(u,k^2)=\cos \psi \mbox{, where } u=\int_0^{\psi} \frac{d\theta}{\sqrt{1-k^2 \sin^2 \theta}}.
\end{equation}
There are also
\begin{equation}
\mbox{sn}(u,k^2)=\sin \psi,
\end{equation}
and
\begin{equation}
\mbox{dn}(u,k^2)=\sqrt{1-k^2 \sin^2 \psi}.
\end{equation}
Some useful identities are the following:
\begin{eqnarray}
\mbox{sn}^2(u,k^2)+\mbox{cn}^2(u,k^2)=1  \\
\mbox{dn}^2(u,k^2)=1-k^2 \mbox{sn}^2(u,k^2)  \\
\frac{d}{du} \mbox{cn}(u,k^2)=-\mbox{sn}(u,k^2)\mbox{dn}(u,k^2).
\end{eqnarray}
Roughly speaking, as we can generalize a circle to a ellipse, Jacobi elliptic functions are a generalization of trigonometric functions. When $k^2 \rightarrow 0$, we have
\begin{eqnarray}
\mbox{sn}(u,k^2)=\sin u, \\
\mbox{cn}(u,k^2)=\cos u, \\
\mbox{dn}(u,k^2)=1.
\end{eqnarray}

\section*{Acknowledgement}
This work is supported by the Ministry of Science and Technology (MOST) of Taiwan under grant number MOST 106-2112-M-167-001.

\end{document}